\def\simgt{\lower.5ex\hbox{$\; \buildrel > \over \sim \;$}}
\def\simlt{\lower.5ex\hbox{$\; \buildrel < \over \sim \;$}}

\documentclass[useAMS,usenatbib]{mn2e}
\usepackage{graphicx}
\usepackage{epstopdf}
\usepackage{multirow}

\usepackage[usenames,dvipsnames]{color}

\newcommand{\msun}{\ensuremath{\, {M}_\odot}}
\newcommand{\Msun}{\ensuremath{\, {M}_\odot}}

\newcommand{\ocen}{$\omega$~Centauri}

\newcommand{\alfafe}{[$\alpha$/Fe]}
\newcommand{\feh}{[Fe/H]}
\newcommand{\cnofe}{[(C+N+O)/Fe]}

\newcommand\aj{{AJ}}
\newcommand\apj{{ApJ}}
\newcommand\apjl{{ApJ}}
\newcommand\aap{{A\&A}}
\newcommand\mnras{{MNRAS}}
\newcommand\nat{{Nature}}

\title[Multiple stellar populations in \ocen.]{The mosaic multiple stellar populations in \ocen : the Horizontal Branch and the Main Sequence}
\author[Tailo et al.]{M. Tailo$^{1,2}$\thanks{E-mail:
mrctailo@gmail.com}, M. Di Criscienzo$^2$, F. D'Antona$^2$, V. Caloi$^{3}$, P. Ventura$^2$\\
$^{1}$Dipartimento di Fisica, Sapienza Università di Roma, Piazzale Aldo Moro 5, I-00185, Roma, Italy.\\
$^{2}$INAF- Osservatorio Astronomico di Roma, via di Frascati 33, I-00040 Monteporzio, Italy \\ 
$^{3}$INAF, IAPS, Roma, via Fosso del Cavaliere 100, I-00133 Roma, Italy\\
}
\begin{document}

\date{Accepted 2016 February. Received 2015 December; in original form}

\pagerange{\pageref{firstpage}--\pageref{lastpage}} \pubyear{XXX}

\maketitle

\label{firstpage}

\begin{abstract}
We interpret the stellar population of \ocen\ by means of a population synthesis analysis, following the most recent observational guidelines for input metallicities, helium and \cnofe\ contents. We deal at the same time with the main sequences, sub-giant and horizontal branch data. The reproduction of the observed color magnitude features is very satisfying and bears interesting hints concerning the evolutionary history of this peculiar stellar ensemble. 
Our main results are: 1) no significant spread in age is required to fit the colour-magnitude diagram. Indeed we can use coeval isochrones for the synthetic populations, and we estimate that the ages fall within a $\sim 0.5$ Gyr time interval; in particular the most metal rich population can be coeval (in the above meaning) with the others, if its stars are very helium--rich (Y$\sim$0.37) and with the observed CNO enhancement (\cnofe\ = + 0.7);  2) a satisfactory fit of the whole HB is obtained, consistent with the choice of the populations providing a good reproduction of the main sequence and sub giant data. 3) the split in magnitude observed in the red HB is well reproduced assuming the presence of two stellar populations in the two different sequences observed: a metal poor population made of  stars evolving from the blue side (luminous branch) and a metal richer one whose stars are in a stage closer to the zero age HB (dimmer branch). This modelization also fits satisfactorily the period and the \feh\ distribution of the RR Lyrae stars.

\end{abstract}


\section{Introduction}
\label{intro}

\begin{figure*}
\centering
\vspace{-2cm}
\hspace{1cm}
\includegraphics[width=1.9\columnwidth]{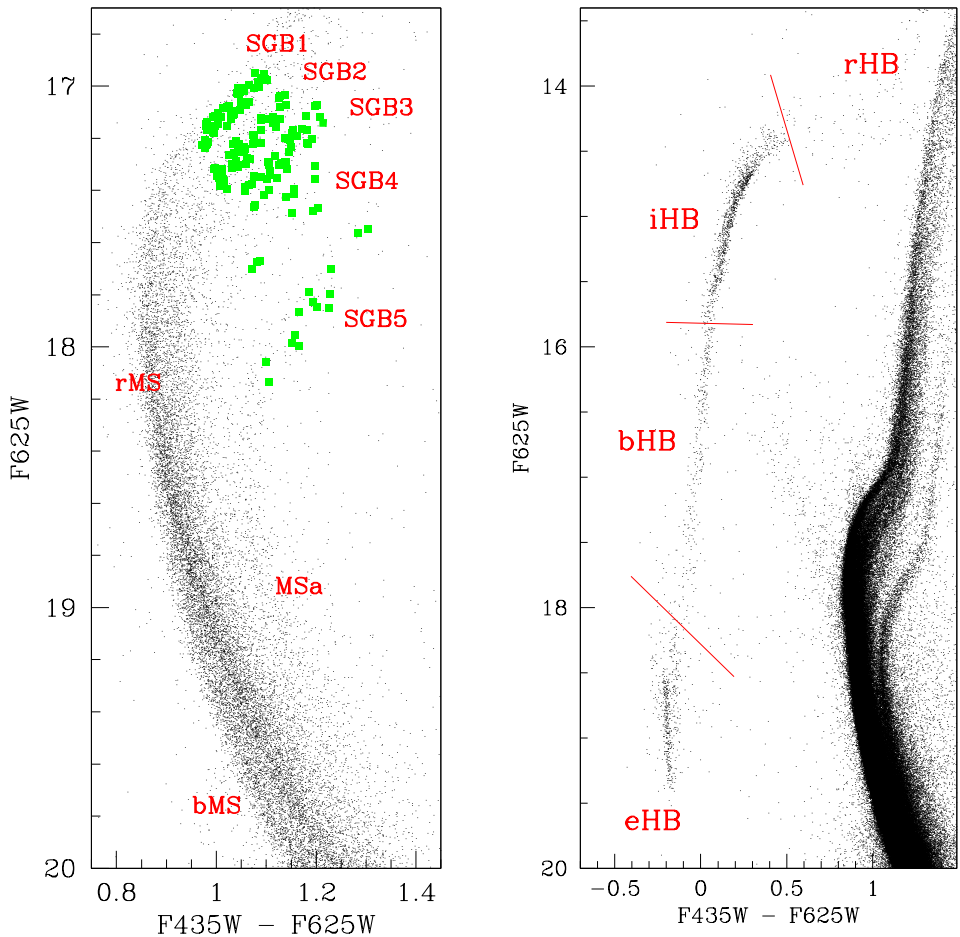} 
\vspace{-2cm}
\caption{\textit{Left panel:} a reduced sample from AVM10 in the F435W and F625W HST bands. Green dots represents the location of the stars observed spectroscopically by V14. Using a combination of these data we identify five sub giant branch (SGB) and three Main Sequence (MS), as labelled in the figure. \textit{Right panel:} the full sample from AVM10 where we label the various section of the Horizontal Branch: the red, intermediate and blue Horizontal Branch (rHB, iHB, bHB respectively) and the  extreme Horizontal Branch (eHB), which also  contains a large population of blue hook stars (BHk).}
\label{PIC_label}
\end{figure*}

Among Globular Clusters, \ocen\ certainly represent the most complex system. Its dynamical and chemical properties led researchers to consider it the surviving nucleus of an ancient dwarf galaxy, captured and disrupted several Gyr ago by the gravitational field of the Galaxy  \citep[see e.g.][]{dinescu1999a,gnedin2002,bekki2006,bellazini2008,marconi2014}. Nevertheless, its chemical evolution has signatures which resemble those of multiple populations in standard Globular Clusters. In fact, while the cluster stars can be subdivided into groups of metallicity spanning a factor 30 or more  \citep[$-2.2 \le $\feh $ \le -0.6$,][]{JP10, marino2011,villanova2014}, inside each of these groups light elements abundances show variations as in mono--metallic clusters, i.e. each has its own O--Na anti correlation \citep{JP10, marino2011}, making not trivial to understand the system chemical evolution and whether this object is more like a globular cluster or a dwarf spheroidal. Recent chemical modelling based on a linear chemical evolution path for each element cannot describe the complex situation  and does not consider the O-Na anti correlations \citep[e.g.][and references therein]{romano2010}.

The photometric data add information on the complexity of \ocen\ star formation episodes. The presence of an extended horizontal branch (HB)  including a populated Blue Hook, the multiple red giant (RGB) and sub-giant branches (SGB), where at least  five stellar populations are distinguishable \citep{villanova2007,bellini2010,villanova2014}, are the most evident signatures. The discovery that the main sequence (MS) of \ocen\ splits into two distinct parallel MS, together with the spectroscopic evidence that their different values of metallicity imply large helium content difference \citep{bedin2004,piotto2005} is another important ingredient which links \ocen\ to simpler (monometallic) clusters, like NGC~2808, which also has a high helium MS \citep{norris2004} and a similarly extended HB morphology \citep{dantona2005, piotto2007}. 

A further photometrically impressive feature of the CM diagram in \ocen\ is the reddest sequence, MSa (see Fig. \ref{PIC_label}), populated by stars having the largest metallicities. They define the lowest luminosity and reddest turn-off. Isochrone fitting indicates that this sequence is populated by high helium stars \citep[Y $> 0.35$,][]{joo2013}. This is in agreement with the suggestion by \cite{dantona2011}, who hypothesized that this metal--rich component of stars, showing direct O--Na correlation, formed from the ejecta of a late generation AGBs. Small or no dilution is expected in this case, because pristine gas was exhausted by the previous episodes of star formation or expelled from the structure \citep{dercole2008} and re-accretion of pristine gas did not occur.

The main question is whether \ocen\ is to be regarded more as an `extreme' GC, whose more complex chemical evolution is merely due to its larger initial mass, or as an intermediate case between GCs and dwarf galaxies, e.g. the remnant nucleus of a satellite galaxy which was destroyed when merging with the Milky Way \citep{norris1997,lee1999}, or a former nuclear star cluster \citep[e.g.][]{boeker2008,dacosta2015}. Dating the different populations of this system is indeed a first step to choose among these possibilities, as GCs star formation is generally believed to be contained in a time span of the order of 10$^8$\,yr \citep{dercole2008}, while much longer time-scales are possible in the other systems \citep{georgiev2014}.

Dating the different populations from the different  CM diagram location of the turnoff or of the sub giant branch is the final aim of this work. The most critical point is to determine the relative age of the MSa with respect to the main (metal poorer) populations. The first studies, until 10 years ago, gave disparate results. In some cases, the MSa is found to be {\it older} than the the main populations by 2--4\,Gyr \citep{ferraro2004,freyhammer2005}, while \cite{sollima2005} discuss that the MSa can be so helium rich to be coeval with the others, and \cite{lee2005} find the MSa {\it younger} by $\sim$1.5\,Gyr, using similar hypothesis concerning helium enhancement. Many other dating works came out, based on disparate methods of age determination \citep[see Table 5 in ][]{stanford2006}, and we do not consider them here. More recently, the latest published work fins this metal richest population 1.7\,Gyr younger \citep{joo2013}. In addition, \cite{villanova2014} find huge age spreads within each population, by dating the subgiants with measured iron content.  Now that more detailed information about the chemistry of the different populations is available \citep{JP10, marino2011,marino2012,villanova2014}  it is worth going back to the timescale of \ocen\ formation.

In this work we built detailed synthetic CMD both for the MSs and sub giant branch(es) and for the whole HB, choosing the different populations in agreement with the most recent spectroscopic direct (for CNO) or indirect (for helium) information, and show that the age difference among all the populations can formally be as low as zero, so this interpretation favors the `fast' formation model of GC-type multiple populations \citep[e.g.][]{dantona2011} compared to the `extended' formation period typical of the nuclear star clusters.

The outline of the present work is the following. \S~2 describes the currently available photometric and spectroscopic data; it also describes the assumptions made when dividing the stars forming \ocen\ into distinct stellar populations. \S~3 and \S~4 describe respectively how the stellar evolution and the synthetic population models have been calculated. \S~5 and \S~6  are dedicated to describe the results of our simulation for the MS and the HB evolutionary phases and compare those results with the available data. Finally in \S~7 we discuss our finding concerning the stellar populations of this GC.

\section{Available Data}
\label{data}

\subsection{Photometric observations}
\label{photo}
We examine the colour magnitude diagram (CMD) of the galactic globular cluster \ocen\ described in \cite{dataA}, hereinafter AVM10, with the tools of stellar population synthesis. The first step in our analysis is to identify the numerous features present in the CMD. Figure \ref{PIC_label} represents two views of the AVM10 HST data in the F435W and F625W ACS/WFC bands. The left panel of Figure \ref{PIC_label} shows about 50000 stars from the sample, chosen from the inner part of the cluster, with the aim to show the multiple MSs and SGBs observed. We over plot with green squares the location of the stars spectroscopically  observed by \cite{villanova2014}, hereinafter V14.\\
Combining these two dataset, we identify three main sequences and five sub giant branches, as labelled in the figure. In naming the three main sequences we follow the common nomenclature \citep{bedin2004, piotto2005,dataA,bellini2010,joo2013}. The bluest MS of the three will be named bMS, the central and most populated one rMS, the reddest one is the MSa. The names of the SGBs follow the same guidelines from other authors (e.g. V14). The most luminous one will be dubbed SGB1 and faintest one SGB5, while the others follow a progressive numeration from top to bottom. In the right panel of Figure \ref{PIC_label} we report the full set of HB data from AVM10. We divide this structure in four section: red, intermediate, blue HB (rHB, iHB, bHB, respectively) and an extreme HB (eHB), which contains a largely populated Blue Hook (BHk), located at the hotter end of the HB. The red lines in the right panel of Figure \ref{PIC_label} give a visual indication on the identification adopted.

\subsection{Spectroscopic observations}
\label{spect}
A large number of populations that differ in light and heavy elements content have been spectroscopically identified, leading  to classify \ocen\ as anomalous globular cluster \citep[and references therein]{JP10,marino2011,villanova2014}, alongside a few other (less extreme) clusters, such as M22 \citep{marino2009}, M2 \citep{yong2014}, and NGC 5286 \citep{marino2015}. \\
 More in detail, the stars populating \ocen\ have a spread in \feh\ of almost 2.5 dex: $-2.3 \le \feh \le -0.4$  \citep{JP10}; $-2 \le \feh \le -0.6$  Marino et al. (2011, MAR11) and  $-2.2 \le \feh \le -0.4$ in V14. In spite of the differences in the distributions observed, these different  authors agree in assigning a metal poor nature to the majority of the stars in \ocen. Also [(C+N+O)/Fe] shows star to star variations. A correlation between the iron content and the C+N+O content is found  \citep{marino2012}, with a maximum of \cnofe$\sim + 0.7$\ for\feh$> -0.75$.  Alongside these two features, this object shows remarkable star to star variations in the light elements content at all \feh\ --  \cite{JP10}, MAR11, \cite{gratton2011}. \\
Direct measurements of helium enhancement, albeit made by few authors and quite difficult to realize, indicate that large variations in the abundance of this particular element are present among the stars of \ocen. \cite{dupree2013} described the analysis of two RGB stars that have similar \feh\ but largely different values of [Na/Fe] and [Al/Fe] values. The results of the analysis point out that these two stars have also very different helium content.  Indeed the authors describe the sodium-rich star having Y$=0.39\div 0.44$ whereas the sodium-poor one has Y$=0.22$. Other authors have realized direct measurements of helium enhancement among the stars of \ocen. \cite{moehler2011} and \cite{bidin2012} analysed a large group of bHB and eHB stars describing how helium changes with the temperature. The peculiar pattern described by the authors strongly suggest that the bHB and the eHB is populated by second generation stars and that the phenomenon of the late helium flash mixing is involved in shaping the blue end of the HB locus.
  
\subsection{Assumptions for the synthetic models}
\label{synt-inputs}

\begin{table*}
\centering 
\begin{tabular}{c c c c c c c c}
\hline
Pop. & Z & \feh  & Y & $N/N_{tot}$ &  Location \\
\hline
\hline
MpI    & $0.0001 \div 0.0006$  & $-2.25 \div -1.8$    & $0.25 \div 0.26$ & $47\%$    & rMS,bMS, SGB1, rHB, iHB \\
MpII   & $0.0006 \div 0.001$   & $-1.8 \div -1.4$     & $0.25 \div0.28$ & $9\%$          & rMS, SGB1,SGB2,rHB , iHB\\
MpIII  &  $0.0006 \div 0.001$         & $-1.8 \div -1.4$          & $0.30 \div 0.37$ & $24\%$      & bMS, SGB2, BHk \\
MiI    & $0.001 \div 0.002$    & $-1.4 \div -1.1$     & $0.25 \div 0.27$ & $6\%$  & rMS, SGB2, SGB3, iHB , bHB \\
MiII   & $0.001 \div 0.002$    & $-1.4 \div -1.1$     & $0.28 \div 0.32$ & $9\%$  & rMS, SGB2, SGB3, iHB , bHB \\
MrI    & $0.002 \div 0.006$    & $-1.1 \div -0.8$     &  $0.30 \div 0.35$ & $3\%$   & MSa, SGB3, SGB4, iHB , bHB \\
MrII   & $> 0.006$           & $> -0.8$           & $0.35 \div 0.40$ & $2\%$      & MSa, SGB5, iHB , bHB\\
\hline
\end{tabular}
\caption{The distribution of global metallicity (Z) and the corresponding value of \feh\ of the stellar populations we identify in \ocen. We also report the \textbf{ assumed} helium content (Y) and the relative number we used as inputs in our simulations. The adopted \alfafe\ values are discussed in the text. Finally the last column reports the location where we locate the stars belonging to that specific population (see \S\ref{MSRES} and \S\ref{HBRES}). Each section of the colour magnitude diagram is defined in the text and represented in Figure \ref{PIC_label}}
\label{TAB_POP}
\end{table*}

Using as starting point the  current \feh\ determinations described in the previous section, we define three ranges of iron content: metal poor (Mp,  $-2.25 \le [Fe/H]\le-1.4$),metal intermediate (Mi, $-1.4 < [Fe/H] \le -1.1$)  and metal rich (Mr, $[Fe/H] \ge -1.1$). The corresponding ranges in Z are:  $0.0001\le Z \le0.001$,  $0.001< Z <0.002$ and  $Z\ge 0.002$. We associate to these populations the \alfafe\ values measured in \cite[Table 1]{gratton2011}: \alfafe\ = +0.2 for Z $<$ 0.002 and \alfafe\ = +0.4 for Z $\ge$ 0.002. To translate the metallicity to \feh\ values we adopted Z$_{\odot}=0.02$. On this basis a large majority of the stellar population in \ocen\ ($\sim$ 80\%) belong to the MP range while the two metals enriched groups (MI and MR) host the rest.\\
The second step is to assign a helium abundance to the stars populating \ocen. Our starting point are the observations describing the [O/Fe] -- [Na/Fe] anti-correlation in \ocen\ discussed in MAR11, so we divide the three populations (Mp, Mi and Mr) into distinct sub-populations using the position of the stars in the [O/Fe]-[Na/Fe] plane, making the assumption that low oxygen, high sodium stars are also helium enhanced, as predicted in the self-enrichment scenario where AGB stars are the main source of pollution  \citep{dercole2008}.\\
We divide the Mp population into three groups. The first one, MpI ($0.0001 < Z < 0.0006$), is assumed to include the majority ($\sim$50\%) of all \ocen\ stars, with either a standard (Y=0.25) helium abundance or very low helium enhancement. MpII includes stars (9\% of the total stars observed) with a slightly higher metal content ($0.0006 < Z \le0.001$) and higher surface helium abundance . The spectroscopic data indicate that this group represents a small fraction of stars. Finally MpIII ($\sim$ 25\% of the total) includes the stars having significantly enhanced helium content (0.35 $\le$ Y $\le$ 0.38) and again $0.0006 < Z \le0.001$. According to our interpretation this is the population whose progeny forms the eHB, including the blue hook \citep{tailo2015}. \\
The Mi population is the most oddly distributed. It has a small fraction of stars (6\%) in the oxygen rich part of the O--Na plane, while the rest is located on the low--oxygen side (9\%). This odd features suggests a bimodal distribution of the helium abundance. We name these two groups MiI and MiII respectively. \\
Finally we divide the metal rich stars into two populations, MrI ($0.002 < Z \le0.006$; 3\%) and MrII ($Z\ge 0.006$; 2\%) and assume for both a highly enhanced Y. These two populations differ in \cnofe\ content, as we assign to the MrII stars the highest enhancement observed. \\ 
Table \ref{TAB_POP} summarizes.all populations identified  from observations which will  be the starting point for our simulations as shown in the next Sections

\section{Evolutionary Models}
\label{models}
The synthetic stellar populations are based on newly calculated evolutionary tracks and isochrones for the evolution of low mass stars of the listed Z and Y contents, evolved through the MS, SGB, RGB and HB phases. A secondary outcome of this work is an extensive collection of stellar evolution models to be used in the field of GC study\footnote{The complete sets of models and isochrones trasformed in HST filters F435W and F625W, will be soon  available on site http://www.mporzio.astro.it/ESTA}.

The models have been calculated using the stellar evolution code ATON 2.0, fully described in \cite{Aton}. Convection has been treated according to the Full Spectrum of Turbulence model \citep{FST}. We used the opacity tables from the OPAL website \citep{OPAL_1}, and for the low temperature opacities we followed \cite{Ferguson}. The conductive opacities by \cite{Potekhin} are adopted. When a variation of C-N-O abundances is introduced, and appropriate OPAL opacities are not available,  we used the tables provided by the AESOPUS 1.0 website \citep{AESOPUS}. The mixtures have Z ranging from $Z=0.0001$ to $Z=0.001$, with $\alpha$-elements ratio set to \alfafe $=+0.2$, and from $Z=0.001$ to $Z=0.008$, with \alfafe $=+0.4$  \citep[this choice has been made following the data of][]{JP10,gratton2011}. The choices assure consistency with the measured values of $[Fe/H]$. We also include two sets of models at $Z=0.006$ and $Z=0.008$ with \alfafe $=+0.4$, corresponding to $[Fe/H] \sim -0.8$ and $[Fe/H] \sim -0.6$, with the C-N-O enhancement \cnofe = +0.7   \citep{marino2012}. Finally, for each Z we compute models for four initial $Y=0.25,0.28,0.35,0.40$. 

For each chemical composition we computed evolutionary tracks with mass in the range 0.35\msun $<$ M $<$ 1.10\msun, spaced with a step of 0.05\msun; the evolution was followed from the pre-MS up to the tip of the RGB. Mass loss is described by Reimers' formula \citep{reimers1975}, with the free parameter entering the formulation set to $\eta = 0.3$, in all the relevant models. Based on the evolutionary tracks, we calculated 1 Gyr spaced isochrones.

To calculate the HB models we used the core mass at the RGB tip, derived from the evolutionary models. The HB tracks were calculated as in \cite{dantona2002}. We consider extra-mixing from the border of the convective core by assuming an exponential decay of convective velocities, with an e-folding distance of 0.02Hp (where Hp is the pressure scale height). This assumption is based on the calibration of core overshooting given in \cite{Aton}. The Hb models where calculated  for various chemical compositions, differing in metallicity, helium and alpha-enhancement. For each chemistry the interval of masses considered ranges from the minimum mass, compatible with the corresponding core mass, up to 1.0\msun. Mass spacing for M $>$ 0.5\msun\ is 0.01\msun; for M $\leq$ 0.50\Msun, the mass step is chosen to build models as close as possible to the core mass. From this point of view the last model of each set is meant to represent the least massive model undergoing a standard, non mixed, helium flash. Each model is evolved from the zero age horizontal branch (ZAHB) to the end of the core helium burning phase.

\begin{table}
\hspace{-0.6cm}
\begin{tabular}{c c c c c c}
\hline
Pop. &  Z & $\sigma_Z$ & Y & $N/N_{tot}$ & $\Delta M_{RGB}$  \\
\hline
\hline
MpI    & $5.0\times10^{-4}$ & $5.0\times10^{-5}$  & 0.25 & $47\%$   & 0.150    \\
MpII   & $7.0\times10^{-4}$ & $7.0\times10^{-5}$   & 0.28 & $9\%$      &  0.190 \\
MpIII  &  $8.0\times10^{-4}$ & $8.0\times10^{-5}$   & 0.37 & $24\%$     &   0.190 \\
MiI     & $1.5\times10^{-3}$ & $3.0\times10^{-4}$   & 0.25 & $6\%$ & 0.210 \\
MiII    & $1.5\times10^{-3}$ & $5.0\times10^{-5}$   & 0.30 & $9\%$ & 0.230 \\
MrI     & $3.5\times10^{-3}$ & $7.0\times10^{-4}$   &  0.33 & $3\%$   &  0.230  \\
MrII    & $7.0\times10^{-3}$ & $1.4\times10^{-3}$   & 0.37 & $2\%$     & 0.260 \\
\hline
\end{tabular}
\caption{The inputs used to calculate the simulation shown in \S\ref{MSRES} and \S\ref{HBRES}, which represents our best results in the description of \ocen\ CMD. The columns report the value of the metallicity (Z), the associated Gaussian spread ($\sigma_Z$), the helium abundance (Y) and the relative numbers of each population.  The last column ($\Delta M_{RGB}$) reports the mass loss from the RBG used when calculating the HB stellar populations. The exact values of these inputs has been obtained compared the simulations with the data (see text). The age used when calculating all the stellar population is 13Gyr. }
\label{TAB_SINT}
\end{table}

\section{Synthetic population models}
\label{synt}

We built the synthetic stellar population models for the MS and the SGB phases from the isochrones database.

We consider the stars populating the CMD by AVM10 data and we divide them into the groups described in \S\ref{synt-inputs}, with metallicity and helium in the intervals listed in Table \ref{TAB_POP}; however, the exact input values are iteratively adjusted by comparing the results from the simulation with the data to optimise the agreement, as described in the next sections. The final inputs for the metallicity, helium abundance and relative number used are listed in Table \ref{TAB_SINT}; therefore, the simulation reported in the next sections represents our best results in the description of \ocen\ CMD. 
To better reproduce the data, the simulation code allows to optionally include intrinsic spreads in different parameters. In our calculations we introduced a Gaussian spread in the metallicity of each population, as suggested by the study of the histograms in \cite{JP10}, MAR11 and V14. For the metal poor populations the adopted spread is the 10\% of the final value of metallicity. However a better agreement with the observations is reached (see next section) if a larger spread in Z is assumed for Mi and Mr populations according the observations of MAR11 and V14.
The exact value of this spread is listed in Table \ref{TAB_SINT}. Thus, for each star in each group the final metallicity is  $Z_{eff}(Z, \sigma_Z$), where Z and $\sigma_Z$ represent the mean value of the Gaussian curve and its root mean square width, respectively. Their respective values are listed in Table \ref{TAB_SINT}. The mass of each stars is then extracted via a power law initial mass function; chosen to obtain the best results in the comparisons shown in \S\ref{MSRES}. The adopted mass function is  $dN/dM \propto M^{-0.70}$, similar to the one used in the case of NGC 6397 \citep{dicrisci2010}. In its current version, the stellar population synthesis code does not include the possibility to simulate binary stars. The exact position of the simulated stars on the theoretical diagram is obtained via an interpolation system which uses the isochrone database itself. With this recipe we obtain a value of effective temperature, mass and luminosity, that, together with the values of  \feh\  and \alfafe, allows us to translate this theoretical HR diagram into an observational CMD.

The HB stellar population synthesis was conducted following \cite{dantona2002,dantona2008} and subsequent works. We counted the stars populating the observed HB and we divided the stars into groups that have the same number fraction as those used for the MS and the SGB. For each group we calculated $M_{\rm{RGB}}$, the mass at the tip of the RGB, from the isochrone database. The mass of the HB star is obtained by subtracting the mass lost from the RGB, $\Delta M $. For the latter we assumed a Gaussian distribution centred at $\Delta M_{RGB} $, with a spread $\delta M$ to be adjusted through comparison with the HB data.  The mass of each star in the HB phase is then $M_{\rm{HB}} = M_{\rm{RGB}} -\Delta M(\Delta M_{RGB},\delta M) $. The values of $\Delta M_{RGB}$ adopted for each population are listed in Table \ref{TAB_SINT}. Via the HB model database constructed, we obtain values of mass, effective temperature and luminosity that are then translated to the observational plane.

All the evolutionary sequences and synthetic populations have been translated from the theoretical Hertzsprung -- Russell (HR) diagram to the observational CMD by using the color -- Teff relation and bolometric corrections derived from the model atmospheres by \cite{castelli2004}. For HB stars with $11500K<$Teff$<20000$K a solar value of metallicity is used for these transformations, since radiative levitation increases the iron content in the spectra \citep{grundahl1999, momany2004}. Finally, to be able to compare the absolute magnitude values obtained from our simulation with the data, the adopted values of distance modulus and reddening are: $(m-M)_{F625W}=14.06$ and $E(F435W-F625W)=0.242$.

{The comparison of the final stellar population model of the MS and SGB phase is a three step process. A first comparison is made by eye to see whether the simulated sequences fit the data in order to determine the age of the main population. Here we adjust also the mean value of the metallicity and helium abundance in each group. The second step is to compare the histograms of the rectified sequences. Here we fine tune the the values of number fraction of each population used in the simulations. If the match between the two histograms is satisfactory within $1 \sigma$ of the Poisson distribution, we proceed to the last step in this comparison procedure. We realize a residual Hess diagram subtracting two homogeneous diagram realized for the data and the simulation. The comparison of the HB simulation with the corresponding sample of the data is a two step process. First an initial analysis is made by eye comparing the position of each group of stars along the HB. In this phase an initial value for  $\Delta M_{RGB}$ is assumed. Comparing the histogram of both the data and the simulation, the value is then refined until a satisfactory match, within $1\sigma$ of the Poisson distribution, is reached.

\begin{figure*}
\vspace{-2.3cm}
\includegraphics[width=1.85\columnwidth]{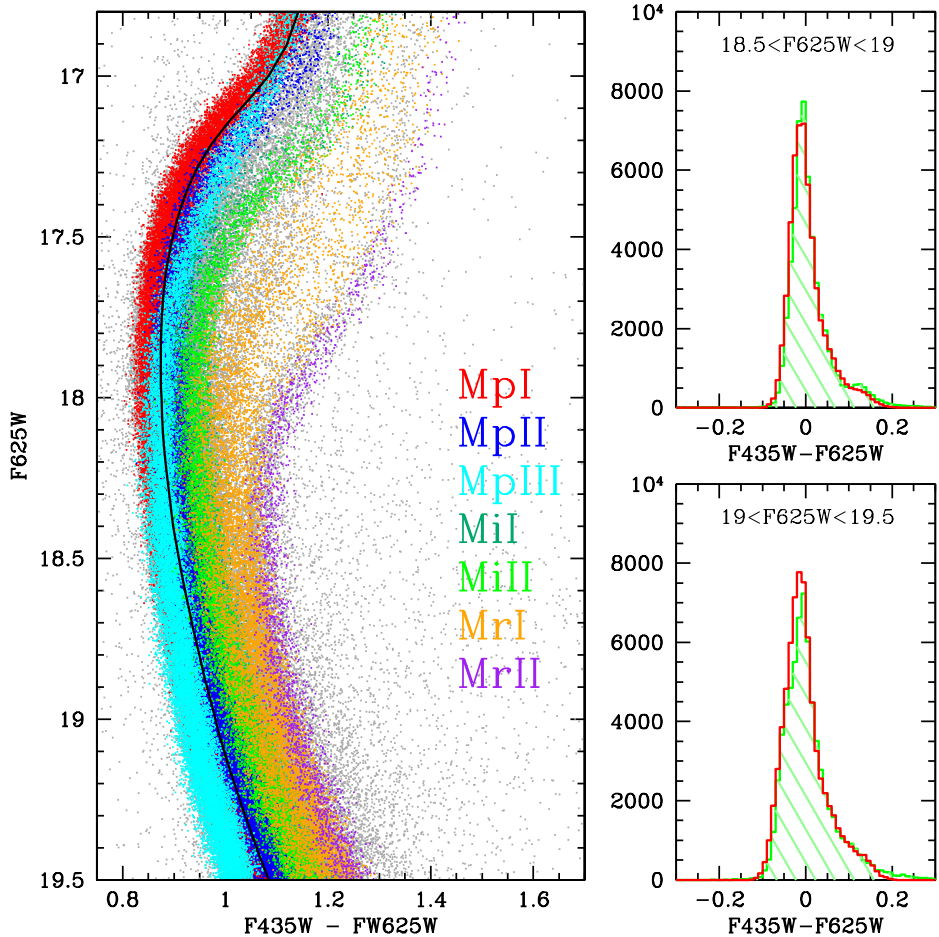} 
\hspace{-1cm}
\vspace{-2cm}
\caption{\textit{Left panel:} The comparison of our best simulation with the data from AVM10 in the F435W and F625W bands (grey dots).  In the figure we represent the upper part of the MS and the SGBs. Each of the populations, identified in \S \ref{synt-inputs} (Table \ref{TAB_POP}), is colour coded in the picture to allow better identification. The simulated populations are all coeval and have an age of 13Gyr.  \textit{Right panels:} The histograms compare the simulation presented in the left panel with the data in the indicated magnitude bins. The sequences have been rectified with the fiducial line represented in the left panel (solid, black). The two profiles refers to data (green, shaded) and the simulated points (red). The adopted distance module and the reddening value are $(m-M)_{F625W}=14.06$ and $E(F435W-F625W)=0.242$. }
\label{PIC_MS}
\end{figure*}

\begin{figure*}
\hspace{-0.5cm}
\vspace{-0.5cm}
\includegraphics[width=1.85\columnwidth]{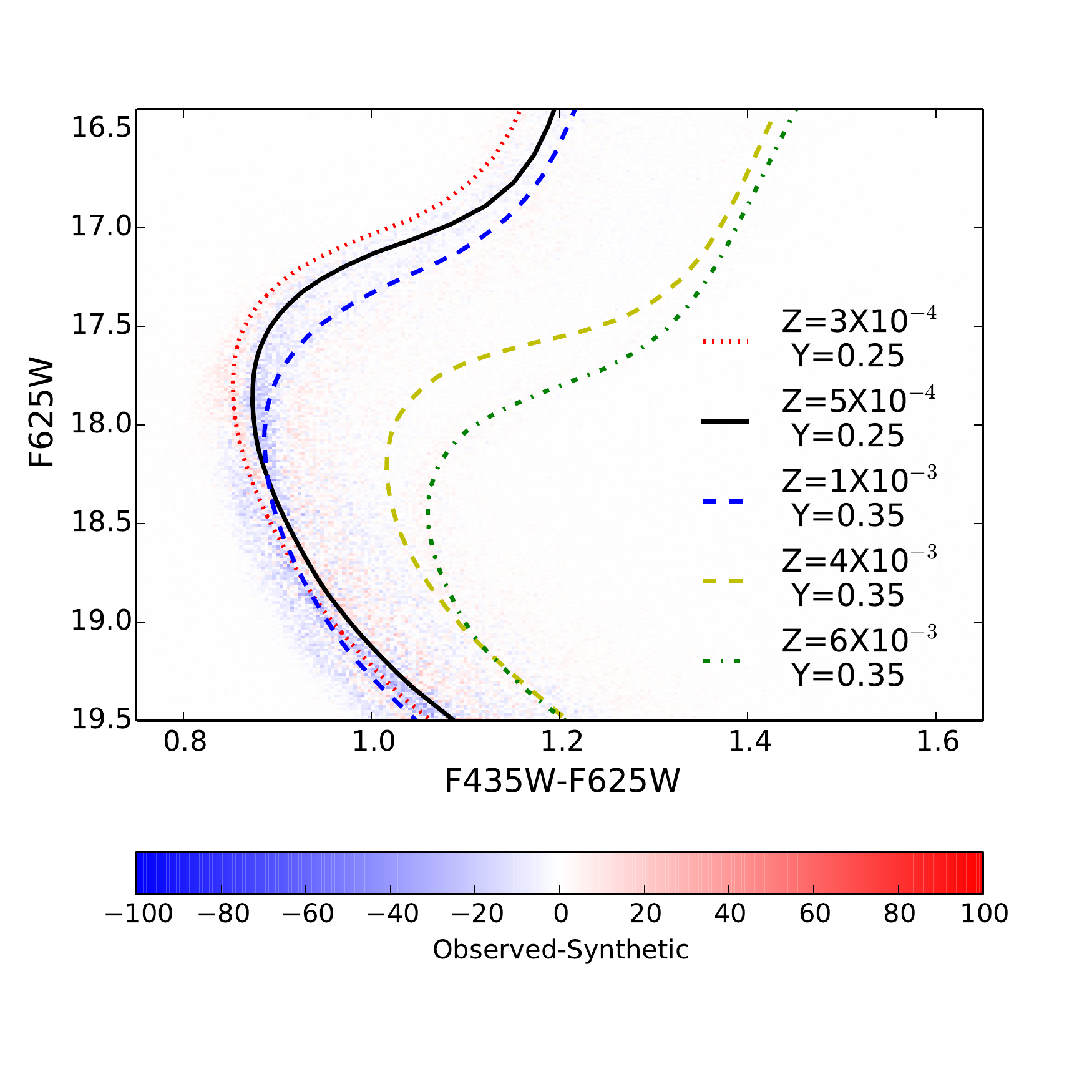} 
\vspace{-1cm}
\caption{ The residual Hess diagram between the observed data and the simulation obtained with the inputs listed in Table \ref{TAB_SINT}. The two homogeneous diagrams, from which the residual have been calculated, are obtained using a $230\times 230$ bin grid of the observational and synthetic CMD. Together with the results of we plot a collection of isochrones from our database, which have been used to construct the simulations. The mixtures used to calculate them is labelled in the figure. It is worth noting that these isochrones have all the same age (13Gyr).}
\label{PIC_HESS}
\end{figure*}

\begin{figure*}
\centering
\includegraphics[width=1.85\columnwidth]{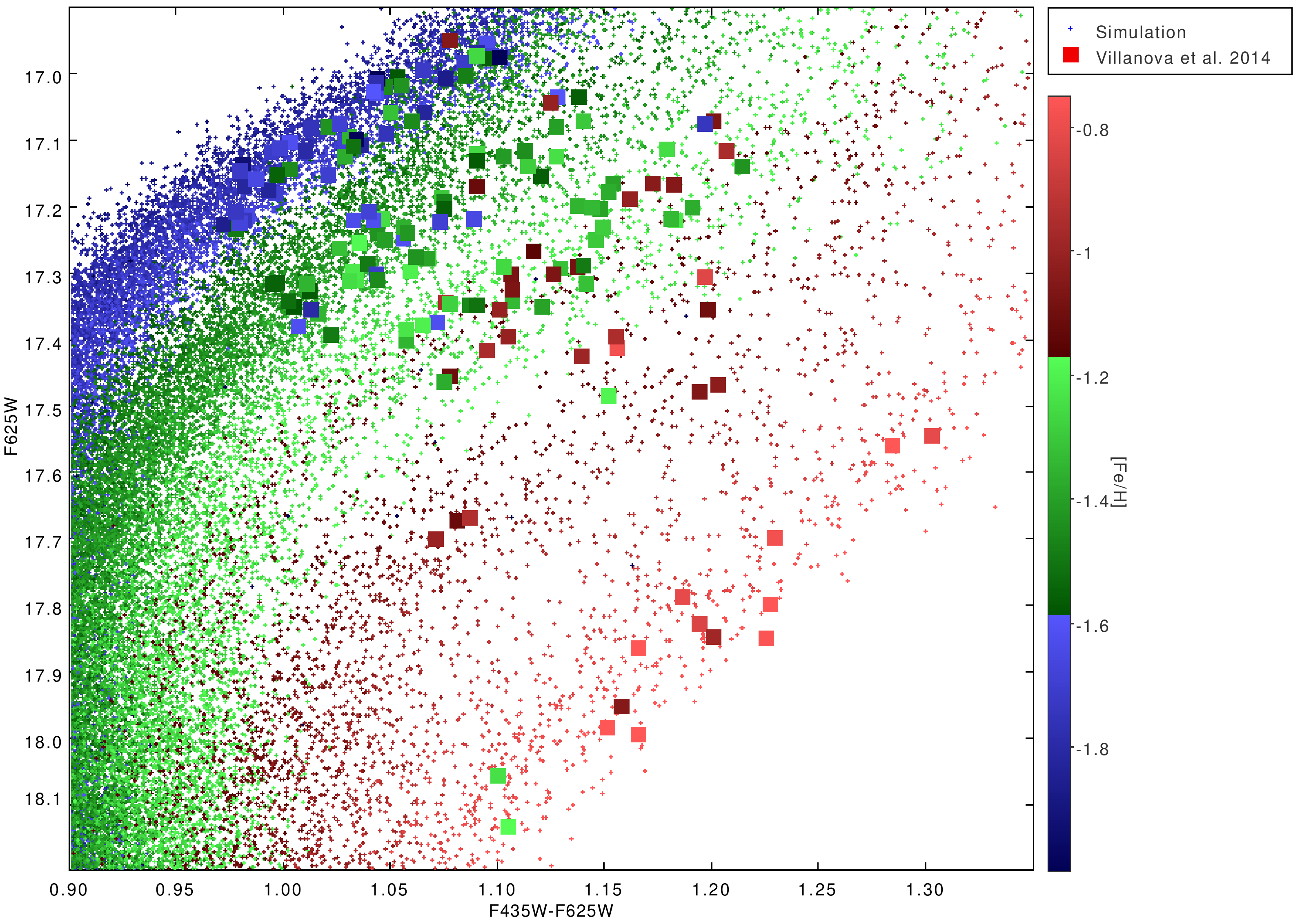} 
\caption{A comparison between our simulation (small dots) and the data from V14 (large squares).The colour coding indicates the \feh\ distribution of the simulated points and of the observed data and provides a way to compare them. The detail of the composition of each SGB we identify is reported in the text. In spite of the coevality of all populations, we find a good agreement between the metallicity in the simulation and the observed \feh\ values in V14.}
\label{PIC_SGB}
\end{figure*}
\section{Main Sequence and Sub-giant branch results}
\label{MSRES}

The comparison of our best simulation with the data  for the upper MS and SGB region is  reported in Fig. \ref{PIC_MS}. The left panel shows the Color Magnitude Diagram where the data from AVM10 are indicated with grey dots. The different populations in our simulation  reported  in Table \ref{TAB_POP} are colour coded  to allow better readability.

As generally recognized, we interpret the rMS as the main stellar population of this cluster. In our simulation the rMS is mainly composed by stars belonging to MpI and MpII, with contaminations from Mi, since the two families overlap in the CMD. Thus the rMS hosts the majority of this cluster's stars, about 55\% (Fig.~\ref{PIC_MS}). The stars located in the bMS are assigned mainly to MpIII, and also include some stars from the most metal poor side of MpI, whose main sequences slightly overlap in the bMS. We  estimate that the bMS hosts about 30\% of the total star of the cluster. The most difficult part of this analysis regards the interpretation of the reddest of the three main sequences, the MSa. It results to be mainly composed by stars belonging to MrII, the most metal rich minority, about 2-3\% of MS stars; as described in \S~\ref{spect} and in \S~\ref{synt-inputs} these stars show a large \cnofe\ enhancement. The remaining 15\% of the MS stars are distributed among the Mi and MrI populations. The simulated populations are all coeval, with an age of 13Gyr.

A more quantitative analysis is shown in the two right panels in Fig. \ref{PIC_MS}, reporting the histogram of the labelled section of the MS.  The histograms include data within two different interval of magnitudes below the turn--off; the colors have been `rectified' using the fiducial line drawn in the left panel (solid, black line). The green, shaded profile refers to the data, the red one to the simulation. The comparison can be done up to F625W $\sim$ 19.5, where the data can still be considered complete (see AVM10). We reach an overall, satisfactory agreement between the two profiles. 

In Fig. \ref{PIC_HESS} we show  the residual Hess diagram of the CMD,  which  allows a general comparison between data and simulation. For the comparison we subtract the two homologous diagrams of AVM10 data and simulation. The red and the blue areas in the figure represent an excess of observed or simulated stars, respectively. The presence of light coloured bins throughout the majority of the diagram demonstrates a satisfactory agreement between observations and simulations. In Fig. \ref{PIC_HESS} we also plot some of the isochrones used in our calculations with the intention of stressing  the evidence that the bMS hosts both standard helium, metal poor stars and he-enhanced, metal intermediate stars (MpIII, Y $\sim 0.37$). Furthermore, an overlapping is also evident for the two metal rich, he-enhanced isochrones forming the MSa.

In Fig. \ref{PIC_SGB} we compare the distribution in the CMD of the points from our simulation and the data from V14 (see \S~\ref{photo}) showing that the simulation is not it contrast with the \feh\ measurements. The figure shows that  each of the five SGBs identified is populated by stars belonging to one or more of the populations we divided the stars into, which can be identified with their \feh\ value. SBG1, the most populated one, comes from the metal poor populations, MpI and MpII mainly. SGB2 contains stars of the MpII population. Both SGBs are also affected by contaminations from the MpIII population, given the shape of the sequence. We interpret SGB3 as made up by stars of the Mi families. SGB4,  the most sparse one, contains stars belonging to the MrI family. These SGBs represent the subsequent evolution of the stars that do not fall into the three MSs identified. Being the subsequent evolution of the MSa, the interpretation of the SGB5 is the most straightforward: it is composed by stars belonging to the MrII population, with enhanced C+N+O surface abundance.

Not only we have reached a good reproduction of the data without introducing a large age spread among different populations; but the corresponding SGBs are consistent with the metallicities found from spectroscopy. It is worth noting that including helium and \cnofe\ enhancement in our simulations has allowed a good representation of the observations.

We adopted stellar populations that are rigorously coeval for convenience and ease of use.
We also checked that the relative age determination of the populations, {\it for fixed chemistry, as listed in Table \ref{TAB_SINT}} can not be changed by more than 500\,Myr to preserve the quality of the fits.
We remark that we are not able to assess the true age spread among the populations of \ocen, as this strongly depends on the chemistry assumed for these populations, namely [Fe/H], Y and \cnofe. Different choices may lead again to larger age differences.  The important result here is that there are combinations of chemical composition, consistent with the observed spectroscopic values and with the helium contents which can reasonably be assigned on the basis of the oxygen - sodium anti correlations, which allow for coevality of all the populations.

To summarize, we achieved a good description of both the photometric and the spectroscopic data of the upper MS and the SGB. \textit{Comparing  our simulations with AVM10 and V14, we find that it is not necessary to assume a large age spread among the different stellar populations to reproduce the observed data.}

\section{Horizontal branch results}
\label{HBRES}

\begin{figure*}
\vspace{-2.5cm}
\hspace{1cm}
\includegraphics[width=1.9\columnwidth]{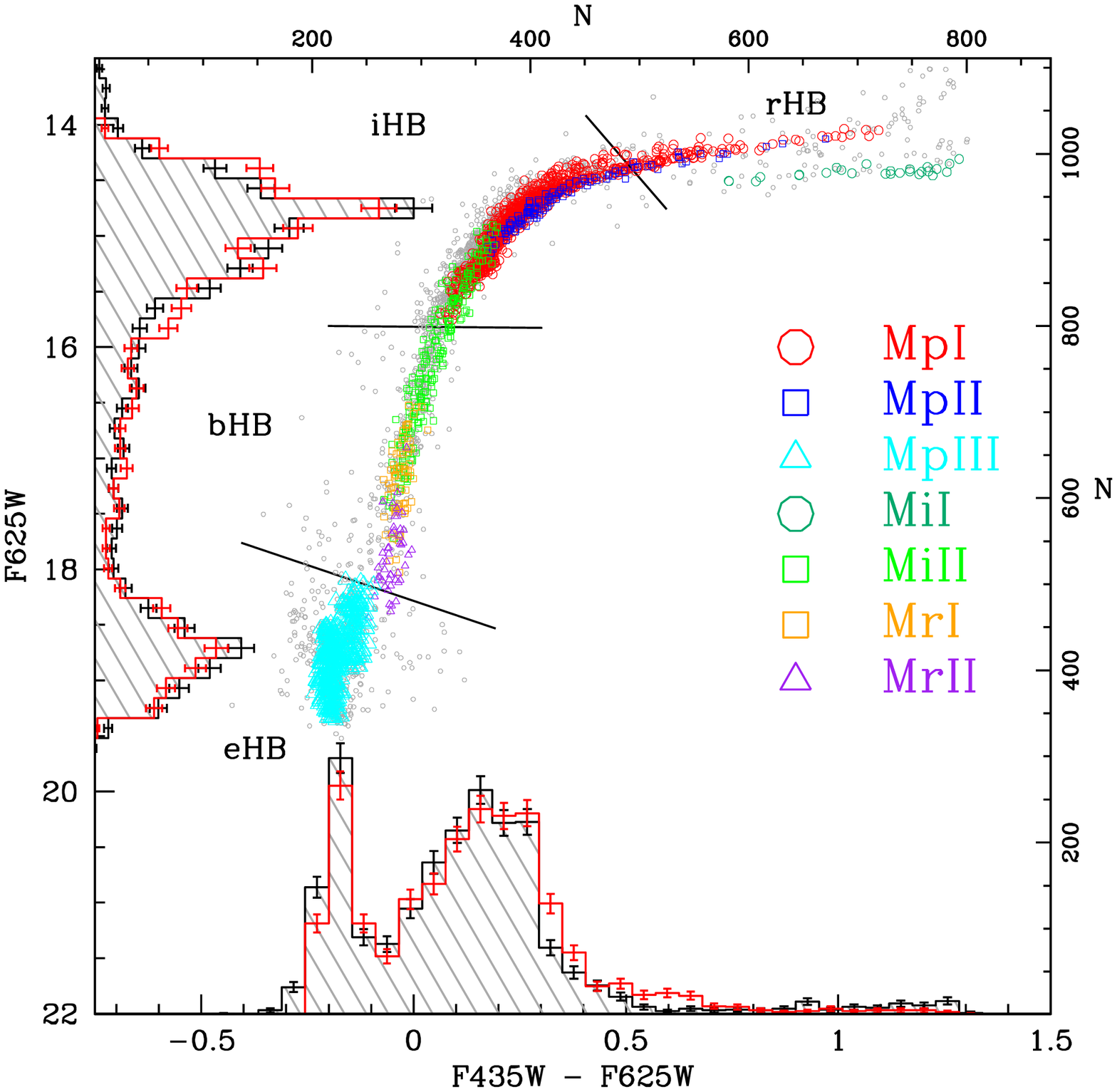} 
\vspace{-2.5cm}
\caption{Comparison of our HB synthetic population model with the sample from AVM10. The colour coding in the figure helps identify the various population we divide the stars. The adopted reddening and distance modulus values are the same of Fig. \ref{PIC_MS}. Description of the populations can be found in Table \ref{TAB_POP}, Table \ref{TAB_SINT} and in \S~\ref{synt-inputs}. We perform a quantitative comparison via the histograms shown in the figure. The black, shaded profile refers to the observed data whereas the red one describe the simulated points. The error bars are defined as 1$\sigma$ of the Poisson distribution. In this picture the simulation of the Blue Hook part of this cluster has been obtained as in Tailo et al. (2015). The black, solid lines report the same division described in \S~\ref{photo} and in Fig.~\ref{PIC_label}}
\label{PIC_HB}
\end{figure*} 

Synthetic HB modeling, based on the prescriptions given in Table 1, was used to understand the stars populating this region of the CMD and as a test of our overall interpretation. The number fractions of the different populations correspond to those used to obtain the CMD shown in Fig. \ref{PIC_MS}. The simulated HB is represented in Fig. \ref{PIC_HB} and, compared with the data of  AVM10 (small, grey circles). The color--coding  in Fig. \ref{PIC_HB} is the same as in Fig. \ref{PIC_MS}, as well as the distance modulus and the reddening adopted.

We base our analysis on the comparison between the number count histograms indicating the magnitude and colour distribution of the observation (black line, grey shaded) and simulation (red line, no shading). The error bars of the histograms are calculated as $1\sigma$ from the Poisson distribution. 

 In agreement with other indications from the literature, we kept the mass loss spread very small \citep[as in the case of M3 and in Fornax GCs,][]{caloi2008,dantona2013}: we set $\delta$M=0.006\msun. This particular value is also suggested by the distribution of the periods of the RR Lyrae stars of \ocen, as described in \S~\ref{RRRES}.  The mass loss for each group has been fine tuned to obtain the most adequate number versus color and magnitude distribution (Fig. \ref{PIC_HB}  and column 6 of Table \ref{TAB_SINT}). The mass loss  increases with both Y and Z, suggesting that  second generation stars are subject to stronger RGB  mass loss than the first generation counterparts  \citep[see also][]{dantona2013,salaris2008, tailo2015}.\\
The rHB stars are located along two separate sequences, which merge into the iHB, where the ``knee'' of the HB is located. The splitting is in agreement with the presence of two stellar populations differing in Z  \citep{sollima2006}. We find that the brighter part hosts the subsequent evolution of the iHB stars (see also Fig \ref{PIC_RR}) thus the stars located here have the same mean mass (see next). On the other hand, the stars populating the fainter sequence are more metal rich and slightly less massive, with a mean mass of 0.64\msun. The MpI, MpII and MiI groups, with helium abundance close to primordial Y, populate the rHB, $\sim$8\% of the total number of HB stars.\\
The iHB hosts most of the HB stars (52\%) and contains MpI and MpII stars, plus contaminations from MiII (Fig.~\ref{PIC_HB}). The evolutionary path of these stars crosses the RR~Lyr instability strip starting from its blue side (see Figure \ref{PIC_RR} and the next section). The mean mass of the stars living in this part of the HB is about 0.65\msun, in agreement with other estimates \citep{cassisi2009}. The helium spread among the iHB stars is $\Delta Y \sim +0.03$ above the primordial helium abundance (see Table \ref{TAB_SINT}).\\ 
The bHB, in the range 15.7 $<$ F625W $<$ 18, includes part of the MiII population and the two Mr groups, all helium rich populations ($\Delta Y \ge 0.05$) and enhanced in \cnofe. The mass of the iHB stars is in the range 0.56--0.47 \msun, for a fraction of about 12\% of the total.\\ 
Finally, as in our previous work  \citep{tailo2015}, we interpret  stars in eHB as the helium burning phase of the most he--rich part of the metal poor population (MpIII in Table \ref{TAB_POP} and Table \ref{TAB_SINT}).  A detailed discussion of the modelling of the blue hook stars  which account for about two thirds of eHB which requires treatment of elements diffusion and rotation, is contained in \cite{tailo2015}. 
  
\subsection{RR Lyrae}
\label{RRRES}

In order to compare the synthetic HB with the observed RR Lyrae period distribution, we determine the blue and the red boundaries  of the instability strip  \citep[equation 3 and 4]{marconi2015} and assign to each star in the instability region a pulsation period calculated using equation 1 in \cite{marconi2015}.\\
Fig. \ref{PIC_RR} shows a magnified view of the HB centered on the RR Lyrae region. As in  Fig. \ref{PIC_MS} each star is colour--coded according to the distinctions used in the previous sections, and  we also plot some of the evolutionary tracks involved in the simulation. The splitting  in magnitude observed  is due to the presence of two stellar populations differing in Z: the  bright  branch includes  metal poor stars belonging to the MpI population, with also some stars from the MpII group; the low luminosity branch is  made up by  the metal richer stars from MiI population. \\
In Fig. \ref{PIC_RRdata} we compare the simulated RR Lyrae with the data by \cite{sollima2006}, in the \feh\ vs $\log$P plane. Filled and empty circles identify  respectively RRab and RRc. The simulated RR~Lyrae stars are shown as filled, red triangles. The periods of the observed  RRc have been converted into fundamental periods using the relation $\log{\rm P}_{\rm F} = \log{\rm P}_{\rm F_0}+0.13$ \citep{dicrisci2004}. The histograms describe the distribution of the observed and simulated periods values for the metal rich component of the sample ($\feh > -1.4$, top) and for the metal poor component ($\feh < -1.4$, bottom ) respectively and  shows the adequacy of the simulation.\\
We remark that the two groups of RR Lyrae differing in metallicity described in \cite{sollima2006}, according our interpretation, are also in different evolutionary stages. The metal richer stars are closer to their ZAHB phase, and the lower metallicity ones are in an advanced stage of their helium burning evolution, crossing the boundary of the instability strip from the blue side, as shown in Fig.6. It is worth noting that  the metal rich sample includes only RRab variables, so this result  supports a possible role of  the evolutionary path through the instability strip in determining the pulsation mode \citep[the hysteresis mechanism, originally proposed by][]{Albada1973}.\\
The  agreement  with  both the RR Lyrae and photometry data is possible only by keeping the mass loss spread very small, as  previous found in other galactic and extragalactic globular clusters \citep[e.g. M3 and the GCs in Fornax dwarf spheroidal;][]{caloi2008,dantona2013}.

\begin{figure}
\centering
\hspace{1.5cm}
\includegraphics[width=1.1\columnwidth]{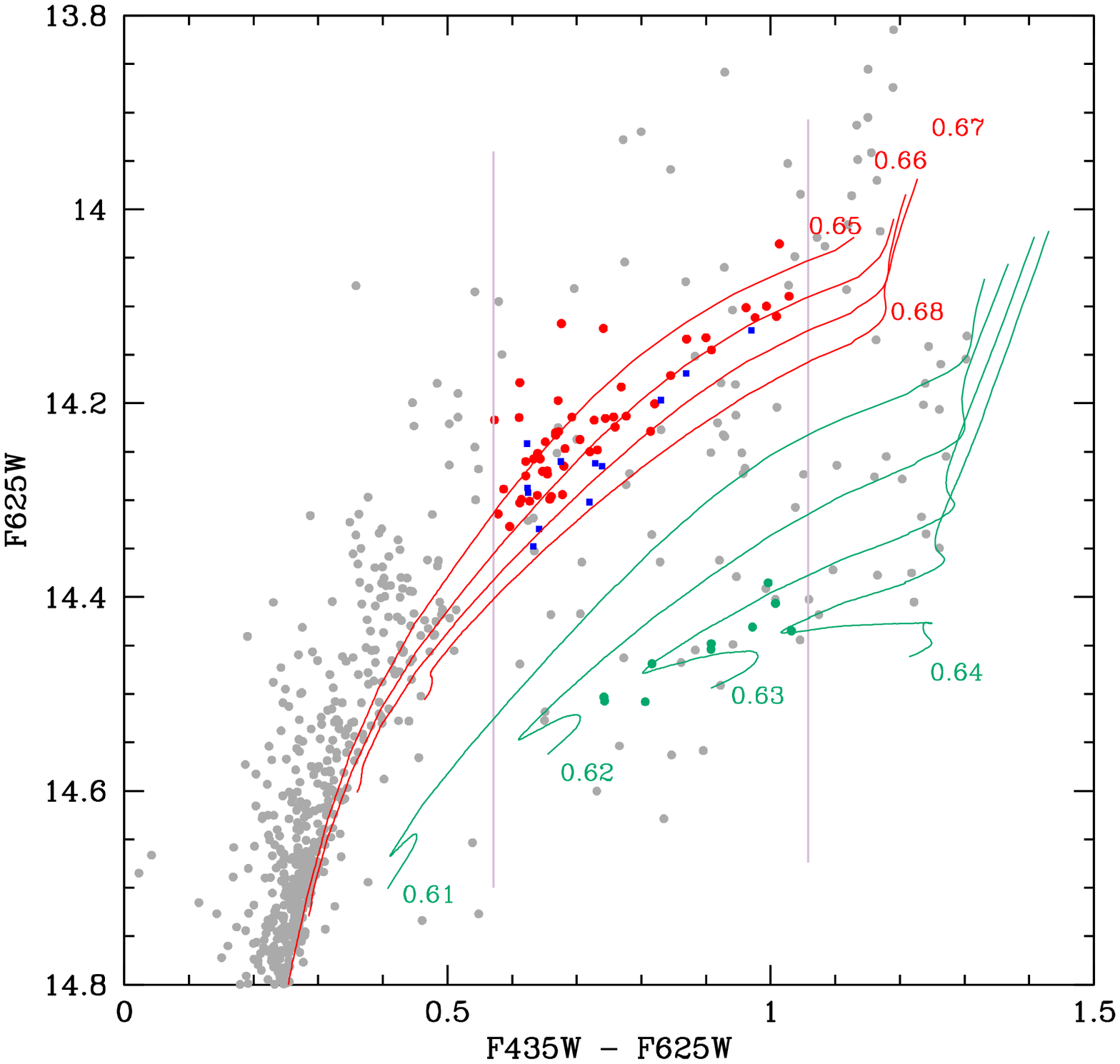} 
\vspace{-1.5cm}
\caption{A magnified view of the instability strip of  the rHB in \ocen\ where are plotted the boundaries of the Instability strip obtained by Marconi et al. (2015) We also plot in the figure the tracks involved in the simulation of these particular stars. The colour coding adopted is the same as Fig. \ref{PIC_HB} to allow an easier identification of the stars family. We remark the fact that the metal poor stars cross the boundary of the instability strip from the blue side and are in an advanced stage of the helium burning phase while the metal intermediate stars are closer to the ZAHB locus of the MiI populations.}
\label{PIC_RR}
\end{figure} 

\begin{figure}
\centering
\hspace{1.5cm}
\includegraphics[width=1.1\columnwidth]{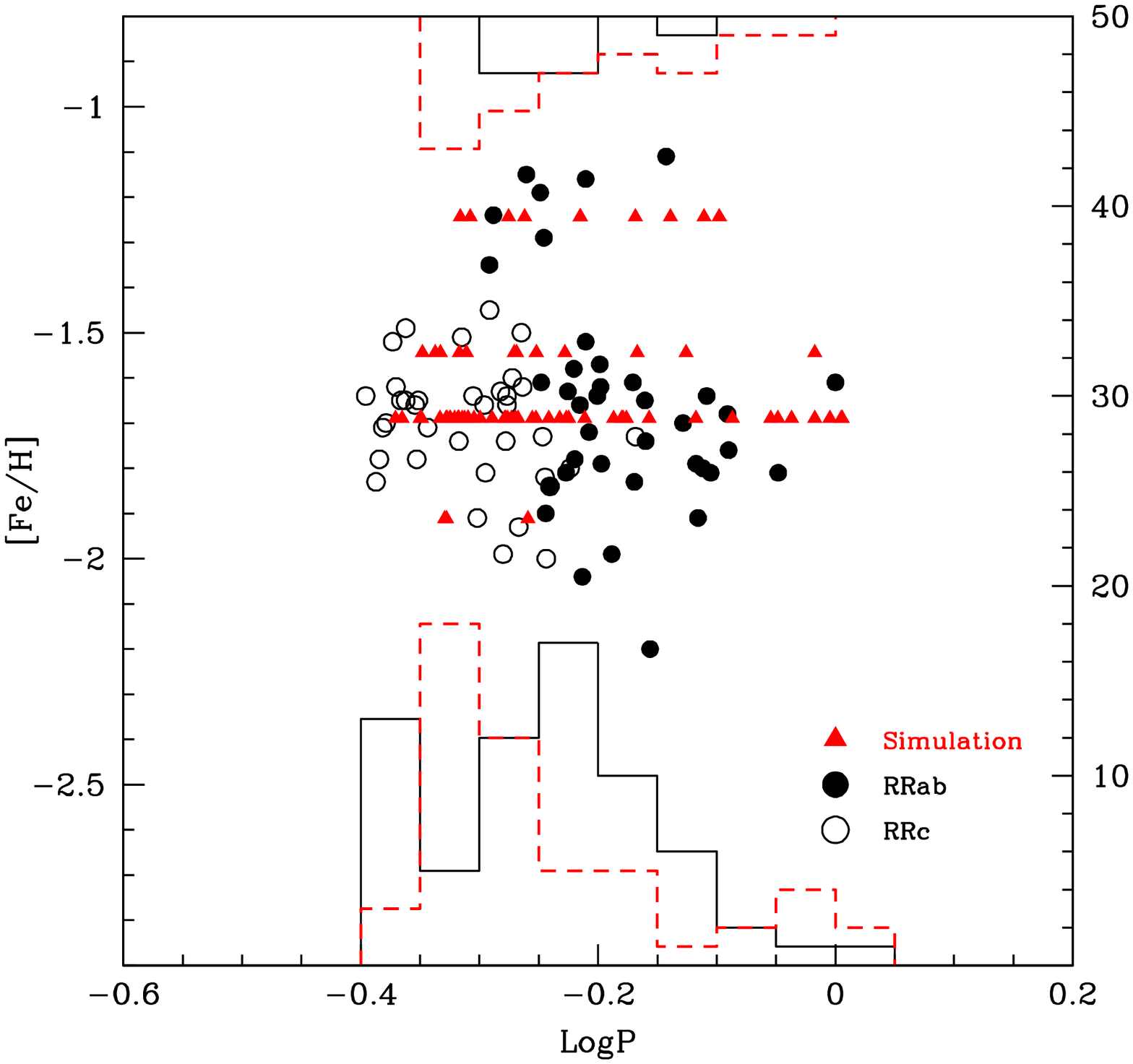} 
\vspace{-1.5cm}
\caption{The comparison of RR Lyrae we find in our best  simulation  r(ed triangles),  with the data from Sollima et al. (2006), circles, in the \feh\ vs log(P) plane. RRc and the RRab, are shown with empty and filled circles respectively. The RRc periods have been fundamentalized (see text). The histograms represent the distribution of log(P) of the variables in the simulation (red profiles) and the data (black profiles). The upper histograms refer to the intermediate metallicity stars ($\feh > -1.4$), while the lower ones to the metal poor  ($\feh < -1.4$) variables. }
\label{PIC_RRdata}
\end{figure} 

\section{Discussion}
\label{conlcusions}
\ocen\ is a complex atypical globular cluster, probably remnant of a complex star formation history, shown by the presence of multiple metallicity groups, and of the typical O--Na anti-correlation within each \feh\ group, implying, for each of them, a variety of helium contents. One of the key problems in the history of \ocen\ is how long  the star formation of all the system components lasted. This implies dating the different populations, reconstructing the CMD properties as well as possible, based on the available spectroscopic and photometric data.\\
For this main aim, we built  synthetic CMDs based on a new stellar evolution models database for the main sequence, red giant branch and horizontal branch, including masses up to 1.1 \msun. The database covers the entire range of \feh\ values measured for \ocen\ in MAR11 and V14, four values of Y for each \feh, and a number of computations including \cnofe\ enhancement for the models in the higher metallicity range, as measured in \cite{marino2012}.  \\
Our best simulation reproduce the large mosaic of stellar populations of \ocen. We list the main results which emerge from our modellization.
\begin{enumerate}
\item For the MS region, we find that the three main MSs of this cluster are composed by a collection of well more than three populations, listed in Table \ref{TAB_POP}. The rMS includes stars belonging to the MpI, MpII and MiI populations; the bMS includes stars belonging to the MpIII, plus the most metal poor part of the MpI. The faintest (or coolest) main sequence, the MSa, includes stars belonging to the MrII population. These latter stars have a significant helium enhancement ($Y \sim 0.37$) and a high \cnofe\ value. 

\item The diversity of stellar population that we identify in the MS reflects itself in an equally diversified mapping of the SGB region. Among the five SGBs identified in the data, the most peculiar is the faintest one, SGB5. Being the direct prolongation of the MSa, it is composed of stars enriched in metal and in helium content and showing a high \cnofe\ value. The peculiar composition of the stars populating this SGB  is crucial to obtain a correct match of the morphology observed, as already explained by  \cite{joo2013}, and is in agreement with the prediction that it is the signature of the last stars formed in the cluster, which are formed from pure AGB ejecta \citep[MAR11,][]{dantona2011}. 

\item The HB, simulated using inputs consistent with the ones used for the MS, and some choice concerning the mass lost on the RGB for each group, reflects the large variety of stellar populations. The rHB results composed of stars belonging to the MpI and MiI population, with few contaminations from MpII. The splitting found in the data of this part of the HB is recognized as a clear signature of the presence of two different metal contents. The iHB hosts most of the HB stars  and it is composed by stars belonging to the MpI, MpII and the MiII populations. The bHB hosts the metal rich component, the MrI and MrII stellar populations. A more detailed model, involving the phenomenon of late helium flash mixing, is required to describe the full photometric observations of the blue hook \citep[and references therein]{tailo2015}. From this point of view the stars located here are the progeny of the bMS, mainly composed of MpIII stars, which has been shown to be more centrally concentrated in the cluster \citep{bellini2009}. 

\item The synthetic HB matches well the observations, although a free parameter (the mass loss in the RGB for each group) must be added. We remark that 1) The groups are chosen to have a narrow mass loss spread (0.006\msun), in agreement with past works related to galactic globular clusters \citep{caloi2005,caloi2008} and extragalactic ones \citep{dantona2013}. 2) The mass loss needed to simulate the iHB, the bHB and, following the results in \citep{tailo2015}, the eHB, has to be increased to reproduce these parts of the HB (see Table \ref{TAB_POP}). Such an increase of mass loss results necessary also from previous modelizations of simpler clusters as NGC2419  \citep{dicrisci2015}, the Fornax dSph clusters \citep{dantona2013}, NGC~1851 \citep{salaris2008}. We suggest that this may be a signature of the dynamic interactions among second generation stars, due to their birth packed into the cluster core \citep{tailo2015}.

\item The comparison with the RR Lyrae variables in \cite{sollima2006} shows that both the period and the \feh\ distributions are reproduced by the simulation. The intermediate metallicity stars in the sample lack a RRc component, while the low metallicity stars present in the data show both RRab and RRc component.  This reflects the behavior of the variables in the CM diagram (Fig. \ref{PIC_RR}) and is associated to the fact that the low metallicity RR~Lyr are evolved stars, while the higher metallicity group includes mostly ZAHB stars. The comparison also highlights the need for a very narrow mass loss spread, as found for M3 variables \citep{caloi2008}.

\item Table \ref{TAB_LIT} reports the results of similar works, i.e. works that use isochrones fitting of the MS and the SGBs, found in the literature giving an estimate of the age spread among the different stellar populations found in \ocen. Our estimate of the age of the main stellar population is consistent, within $1~Gyr$ with most of the work found in the literature, with few exception: most notably we are not in agreement with the estimate from \cite{ferraro2004} and \cite{freyhammer2005}, as the authors conclude that the stellar population forming the MSa belongs to a tidal stream placed behind the cluster itself. Our results are also in agreement with the estimates presented in \cite{lee2005,joo2013}: in this sense both the composition and the number fraction of the stellar population identified, as well as the age of the main population, are coherent.  We however reduce to zero the age difference between the rMS and the MSa. We also find that our chemistry choice, including the observational spread and the errors associated with the spectroscopic analysis, is consistent with an interpretation of coevality among the subgiant stars examined by \cite{villanova2014}.

\item In conclusion, we show that a satisfactory comparison with the most recent spectroscopic and photometric  data does not require a significant age spread to describe all the stellar population of \ocen. We estimate that our fits are consistent with a maximum age spread of $\pm$500\,Myr, if the chemical compositions assigned to the different populations are correct. Thus we answer positively to the question whether there are reasonable choices of composition compatible with an age spread as small as we wish among the populations of  \ocen.

\end{enumerate}

\begin{table}
\centering
\begin{tabular}{c c c c}
\hline
Authors & Age&$\Delta$Age \\
\hline
\cite{ferraro2004} & $15~Gyr$  & $\sim +2~Gyr$\\
\cite{freyhammer2005} &$12\div 13~Gyr$ &$\sim +4~Gyr$\\
\cite{lee2005} &$13~Gyr$ & $-1.5~Gyr$ \\
\cite{sollima2005} &  $16.0~Gyr$ & $ \sim 0\div1~Gyr$\\
\cite{stanford2006} & $13.5~Gyr$ & $- (2 \div 4)~Gyr$\\
\cite{joo2013} & $13.1~Gyr$ & $ - 1.7~Gyr$\\
\cite{villanova2014} &$12.1~Gyr$ & $> \pm 2~Gyr$\\
This work & $13~Gyr$& $0\div -0.5~Gyr$\\
\hline
\end{tabular}
\caption{Summary of isochrones fitting estimates of the maximum age difference among the populations in \ocen. Columns report the age of the main, metal poor, population and the age difference between the most metal rich and the metal poor population. The sign of the figure indicating the age difference stands for an older (+) or younger (-) metal rich population. }
\label{TAB_LIT}
\end{table}

\section*{Acknowledgments}
We warmly thank the referee for its helpful comments in improving the paper. M.T. and F.D'A.  acknowledge support from PRIN INAF 2014 (principal investigator S. Cassisi). M.D.C. acknowledge support from the Observatory of Rome

\end{document}